\documentclass[12pt,a4paper]{kluwer}    % Specifies the document style.

\usepackage{graphicx}
\usepackage{a4wide}

\pubyear{2001}
\firstpage{429}
\spectextone{Astrophysics and Space Science}
\begin{document}                                                                                   

\large

\begin{article}
\begin{opening}         
\title{\large\center\bf{THE FUNDAMENTAL PROPERTIES OF EARLY-TYPE GALAXIES IN THE COMA CLUSTER}} 
\author{STEPHEN A.W. \surname{MOORE}\thanks{Supported by a PPARC studentship. Author email address: s.a.w.moore@durham.ac.uk}}
\author{J.R. \surname{LUCEY}}
\author{H. \surname{KUNTSCHNER}}  
\author{R.L. \surname{DAVIES}}
\institute{Extragalactic Astronomy Group, University of Durham, UK}
\author{M. \surname{COLLESS}}
\institute{Mount Stromlo and Siding Spring Observatories, Australian National University, Australia}
\runningauthor{S.A.W.MOORE ET AL.}
\runningtitle{FUNDAMENTAL PROPERTIES OF EARLY-TYPE GALAXIES IN THE COMA CLUSTER}
%\date{May, 2000}
\dedication{\em{to be published in} {\it{Astrophysics and Space Science}} {\bf{277}} (Suppl.): 429-432, 2001}

\begin{abstract}
We report the results of a high quality spectral study of early-type galaxies within 
the Coma Cluster core.
Stellar population analysis using Lick/IDS indices to break the age/metallicity degeneracy 
are presented, probing their 
formation history and properties. A clear metallicity trend and a dominant single age 
population are found.
\end{abstract}
\end{opening}           

%\small

\section{Introduction}  

A complicated picture has emerged from studies of cluster early-type galaxies,
with measurements of their stellar populations hampered by low quality data and 
by the age/metallicity degeneracy present in broad-band colours. In the core of 
the Coma Cluster, Caldwell {\it{et al.}} (1993) found evidence of a small dispersion in 
the ages of the large majority of early-type galaxies, whilst J{\o}rgensen (1999) found evidence 
of a large spread in age and a small spread in metallicity. In the Fornax 
Cluster, a small age spread and a large metallicity spread was 
found (Kuntschner and Davies, 1998; Kuntschner, 2000). 
These differing results highlight an uncertain understanding of cluster early-type 
galaxy populations. This has important ramifications on studies of the evolutionary 
processes of galaxies in clusters, making it difficult to test hierarchical merging or early 
monolithic collapse models.

Here we present the results of a new study on
 the rich Coma Cluster which aims to accurately measure the ages and metallicities 
of the early-type galaxy population.
Observations were made of the central 1 degree of the cluster with the WHT 4.2m telescope plus the WYFFOS/AUTOFIB2 
multi-fibre spectroscopy instrument. A wide wavelength range $4000\rightarrow{5640}$\,\AA \, was 
studied at high resolution (2\AA \, FHWM) and high signal-to-noise (mean of 50 per \AA).
Many repeat observations over 6 nights were taken to tie down the errors and create a 
homogeneous sample. The Lick/IDS indices H$\beta_G$ (a modified H$\beta$ index proposed by 
J{\o}rgensen, 1999 after Gonz\'{a}lez, 1993) and [MgFe] are used to break the age/metallicity degeneracy.
The data set has a total of 135 galaxies ($m_B$ = 12.6\,--\,18.0).

\section{Data Reduction}
Using any multi-fibre spectroscopy instrument introduces intra-fibre and fibre-to-fibre 
variations in resolution and throughput that necessarily have to be removed before accurate
stellar population analysis can be undertaken. In this study these
effects were mapped extensively and removed.
Standard stars are then used to flux calibrate the spectra. Redshifts and velocity dispersions are
measured using cross-correlation techniques. 
The data is then broadened to the Lick/IDS resolution and the Lick/IDS indices 
measured (see Trager {\it{et al.}}, 1998 for details). Any offsets to the system 
were removed by comparing galaxies in common.

The H$\beta$ index is emission corrected  using a measurement of the OIII strength
(calculated by subtracting a zero emission template from a galaxy spectrum and measuring the 
residual equivalent width) and 
multiplying it by 0.6 to compute the correction (Trager {\it{et al.}}, 2000). A total of 50 
galaxies had 1 sigma evidence of emission, with a median OIII emission of 0.228\AA\, giving 
a median H$\beta$ correction of 0.137\AA\, (corrections are calculated individually for 
each galaxy).

The line index measurement errors were calculated by internal comparison 
during a night and between nights. With the large amount of multiple observations 
with different fibre configurations and
high signal-to-noise data this allows accurate mapping of the  
random and systematic errors.

\section{Comparison with Other Data}

There are 44 galaxies in common with the 
J{\o}rgensen (1999) study which covered a similar area and magnitude 
range, but at a lower signal-to-noise and with multiple data sets. 
The standard deviation between the two studies 
is 0.280\AA\, in H$\beta_G$ and 
0.265\AA\, in [MgFe] --- the principal stellar population analysis line indices 
used in this study. Mehlert {\it{et al.}} (2000) have recently conducted a high
signal-to-noise long slit study of bright early-type Coma Cluster galaxies. 
There are 18 galaxies in common with this study. The long slit data was summed 
up to match the aperture size of this study and compared 
giving a standard deviation of 0.255\AA\, in H$\beta$ (they did not measure the improved index
H$\beta_G$) and 0.103\AA\, in [MgFe].
These comparisons and the internal analysis of the random and systematic errors indicate 
that the study data has a median precision of 0.138\AA\, in H$\beta_G$ and 0.092\AA\, in [MgFe].

\section{Stellar Population Analysis}

\begin{figure}
\tabcapfont
\centerline{%
\begin{tabular}{cc}
\includegraphics[width=7cm]{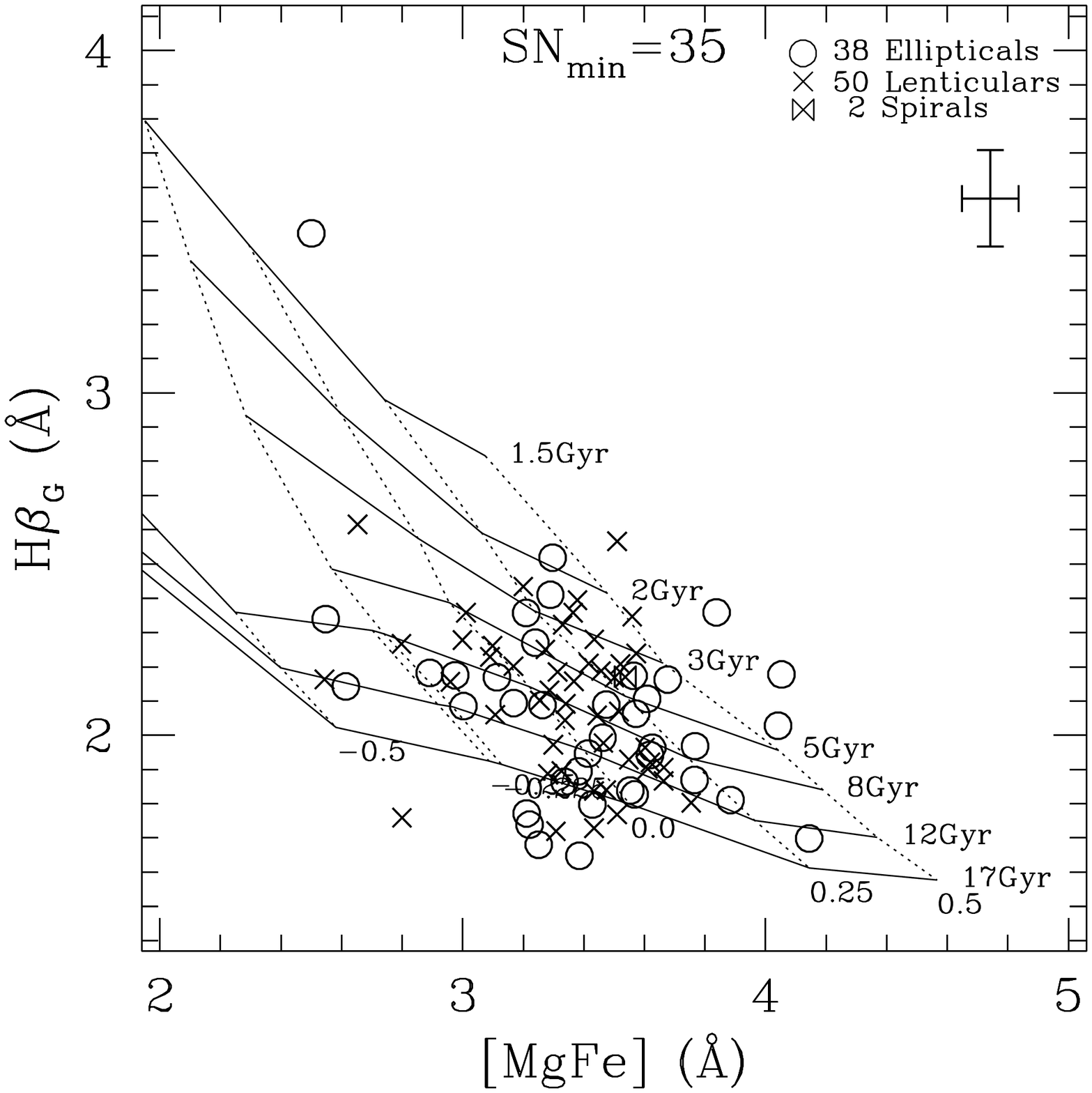} &
\includegraphics[width=7cm]{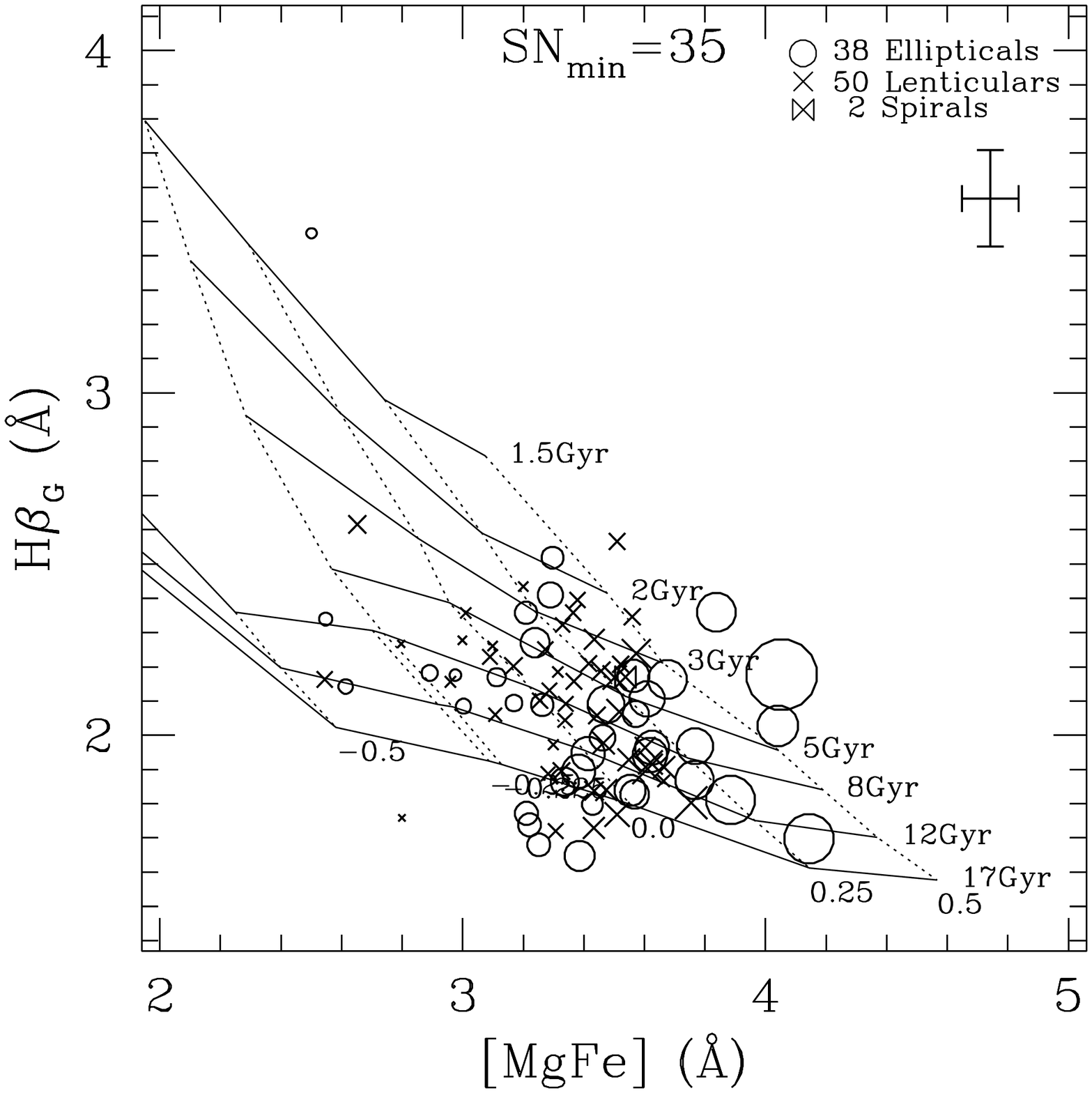} \\
a. & b. \\
\end{tabular}}
\caption{Analysis of Coma Cluster galaxian stellar populations using H$\beta_G$ vs [MgFe]
indicators overlaid on a Worthey (1994) grid. In (a) the symbol size is fixed. In (b) 
the symbol size is scaled to the velocity dispersion of the galaxy 
(larger point size, larger velocity dispersion). Only galaxies with a minimum signal-to-noise of 35 per \AA\, 
are included in the plots.}\label{fig:hb_mgfe}
\end{figure}

This study uses the H$\beta_G$ 
and [MgFe] indices as the principal probe of stellar populations. This 
combination provides
the best compromise to non-solar abundance problems with maximal breaking 
of the age/metallicity degeneracy problem. These indices are superimposed onto 
a Worthey (1994) grid to measure the age and metallicities of the galaxies, probing 
the intra-cluster trends. Figure \ref{fig:hb_mgfe} shows the data, which has been sifted 
to only include data with a minimum signal-to-noise of 35 per \AA.

To test the presence or absence of age/metallicity trends in the cluster we 
used Monte Carlo simulations. These simulations assumed 
a constant age population and performed a density-weighted sampling along a 
given isochrone within the measurement errors. By comparison with the real 
distribution of ages and metallicities within the cluster we can see whether 
a constant age population is supported by the data within the measurement errors.

\begin{figure}
\centerline{\includegraphics[width=9cm]{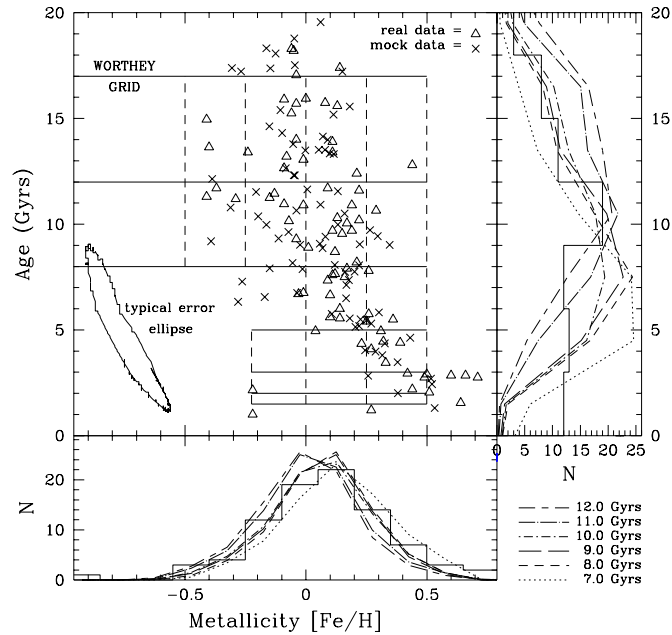}}
\caption{Results of Monte Carlo simulations probing the presence of a dominant single 
age early-type galaxy population (see text).}{\label{fig:monte}}
\end{figure}

The results of these simulations are presented in Figure \ref{fig:monte}. This figure shows 
a plot of age vs metallicity for the real data (triangles) and Monte 
Carlo simulation data for a 9\,Gyr isochrone (crosses) compared to a Worthey (1994) grid. 
To the right and at the bottom of the plot is a histogram of the real data
Overlaid on these histograms are lines showing the results 
from a number of simulations along different isochrones. Comparison of these 
lines to the real age and metallicity distributions shows clearly that a 
dominant luminosity-weighted single age population of 
approximately 9\,Gyrs is supported within the measurement errors. 
A clear metallicity trend, from low to high metallicities, is also seen. This 
metallicity trend is unaffected by differing constant age hypotheses.

\section{Conclusions}

The early-type Coma Cluster data set in this study
is homogeneous, self-consistent, has high signal-to-noise with well characterised 
errors. This has allowed a new unbiased assessment of the 
Coma Cluster intrinsic properties, without any need to combine multiple data 
sets with systematic errors. A stellar population analysis using the indices 
H$\beta_G$ and [MgFe] overlaid on Worthey (1994) grids has shown:

\begin{itemize}
\item{there is a clear metallicity trend;}
\item{the data is consistent with a dominant luminosity-weighted single age 
early-type galaxy population.}
\end{itemize}

\end{article}
\end{document}